\documentclass[10pt,a4paper,twocolumn,english,prl,showpacs,superscriptaddress]{revtex4-1}
\usepackage{mathpazo}

\usepackage[T1]{fontenc}
\usepackage[utf8]{inputenc}
\usepackage{fancyhdr}
\usepackage{color}
\usepackage{babel}
\usepackage{float}
\usepackage{units}
\usepackage{amsmath}
\usepackage{amssymb}
\usepackage{graphicx}
\usepackage[unicode=true,pdfusetitle,
 bookmarks=true,bookmarksnumbered=false,bookmarksopen=false,
 breaklinks=false,pdfborder={0 0 1},backref=false,colorlinks=true]
 {hyperref}
\hypersetup{
 linkcolor=red,citecolor=blue,filecolor=blue,urlcolor=blue}

\makeatletter

\usepackage{babel}
\usepackage{babel}
\usepackage{calrsfs}

\makeatother

\begin{document}
\title{Reservoir engineering with arbitrary temperatures for spin systems
and quantum thermal machine with maximum efficiency}
\author{{\normalsize{}Taysa M. Mendonça}}
\address{Departamento de Física, Universidade Federal de São Carlos, 13565-905,
São Carlos, São Paulo, Brazil}
\author{{\normalsize{}Alexandre M. Souza}}
\address{Centro Brasileiro de Pesquisas Físicas, 22290-180, Rio de Janeiro,
Rio de Janeiro, Brazil}
\author{{\normalsize{}Rogério J. de Assis}}
\author{{\normalsize{}Norton G. de Almeida}}
\address{Instituto de Física, Universidade Federal de Goiás, 74001-970, Goiânia,
Goiás, Brazil}
\author{{\normalsize{}Roberto S. Sarthour}}
\author{{\normalsize{}Ivan S. Oliveira}}
\address{Centro Brasileiro de Pesquisas Físicas, 22290-180, Rio de Janeiro,
Rio de Janeiro, Brazil}
\author{{\normalsize{}Celso J. Villas-Boas}}
\address{Departamento de Física, Universidade Federal de São Carlos, 13565-905,
São Carlos, São Paulo, Brazil}
\begin{abstract}
Reservoir engineering is an important tool for quantum information
science and quantum thermodynamics since it allows for preparing and/or
protecting special quantum states of single or multipartite systems
or to investigate fundamental questions of the thermodynamics as quantum
thermal machines and their efficiencies. Here we employ this technique
to engineer reservoirs with arbitrary (effective) negative and positive
temperatures for a single spin system. To this end, we firstly engineer
an appropriate interaction between a qubit system, a $^{13}C$ nuclear
spin, to a fermionic reservoir, in our case a large number of $^{1}H$
nuclear spins that acts as the spins bath. This carbon-hydrogen structure
is present in a polycrystalline adamantane, which was used in our
experimental setup. The required interaction engineering is achieved
by applying a specific sequence of radio-frequency pulses using Nuclear
Magnetic Resonance (NMR), while the temperature of the bath can be
controlled by appropriate preparation of the initial $^{1}H$ nuclear
spin state, being the predicted results in very good agreement with
the experimental data. As an application we implemented a single qubit
quantum thermal machine which operates at a single reservoir at effective
negative temperature whose efficiency is always 100\%, independent
of the unitary transformation performed on the qubit system, as long
as it changes the qubit state.
\end{abstract}
\maketitle
The study of quantum thermodynamics has been growing in recent years
since it has allowed a deeper understanding of the basics laws of
thermodynamics and its limitations that appear when quantum effects
are taken into account \citep{Gemmer2004}. In this context, quantum
thermal machines, which employ quantum systems as the working medium,
have attracted great interest from physicists since it allows investigating
the fundamental limits of thermal machine efficiencies \citep{Alicki1979,Quan2007}.
In a recent work we have shown, both theoretically and experimentally,
that quantum thermal machines working with one of the reservoirs at
an effective negative temperature \citep{Purcell1951,Ramsey1956,Carr2013,Braun2013}
present counterintuitive behaviors as higher efficiency when performing
non-adiabatic cycles \citep{deAssis2019}, contrary to the usual behavior
of classic thermal machines which provide their maximum efficiency
only in strictly adiabatic processes. In \citep{deAssis2019} the
temperature of the working medium is simulated by properly preparing
the qubit in a thermal state. So, the natural question is: can we
indeed prepare a real reservoir with arbitrary temperature, even effective
negative one, for NMR systems? By employing reservoir engineering
techniques \citep{Poyatos1996,Myatt2000}, here we show, experimentally,
that this is indeed possible. The idea behind reservoir engineering
relies on the manipulation of the system-environment interaction in
order to drive the dynamics of the system to the desired state \citep{Poyatos1996}.
This technique was already successfully applied, for instance, to
investigate the decoherence of motional superposition states a trapped
ion coupled to engineered reservoirs \citep{Myatt2000} and to engineer
vacuum squeezed reservoir for two-level atoms \citep{Murch2013}.
There are also theoretical proposals to apply reservoir engineering
techniques for protecting arbitrary superposition of motional states
of trapped ions \citep{Carvalho2001}, to engineer steady squeezed
states for single \citep{Werlang2008b} or two harmonic modes \citep{Pielawa2007,Werlang2008},
to perform decoherence-free rotations in two-level atoms \citep{Prado2009},
to implement quantum computation and quantum-state engineering driven
by dissipation \citep{Verstraete2009}, and to engineer two collective
spin ensembles individually coupled to the same reservoir \citep{Hama2018}.
Reservoir engineering was also already applied in spin systems using
NMR technique, e.g., to build a time-dependent bath \citep{Suter2011}
or an adjustable system-bath coupling strength \citep{Suter2007}.
Following these ideas, here we show how to engineer reservoirs at
arbitrary temperatures for a qubit spin system, even at effective
negative ones \citep{Purcell1951,Ramsey1956,Carr2013,Braun2013} which
allows for intriguing phenomena in thermal machines \citep{deAssis2019}.
As a special application of reservoirs at effective negative temperatures,
here we implement a new single qubit quantum thermal machine, which
presents the advantage of employing a single reservoir and allowing
for maximum efficiency ($100\%$), independent of the process we carry
out, as long as the final qubit state is different from its initial
one.

\begin{figure}[bh]
\raggedright{}\includegraphics[width=0.8\columnwidth]{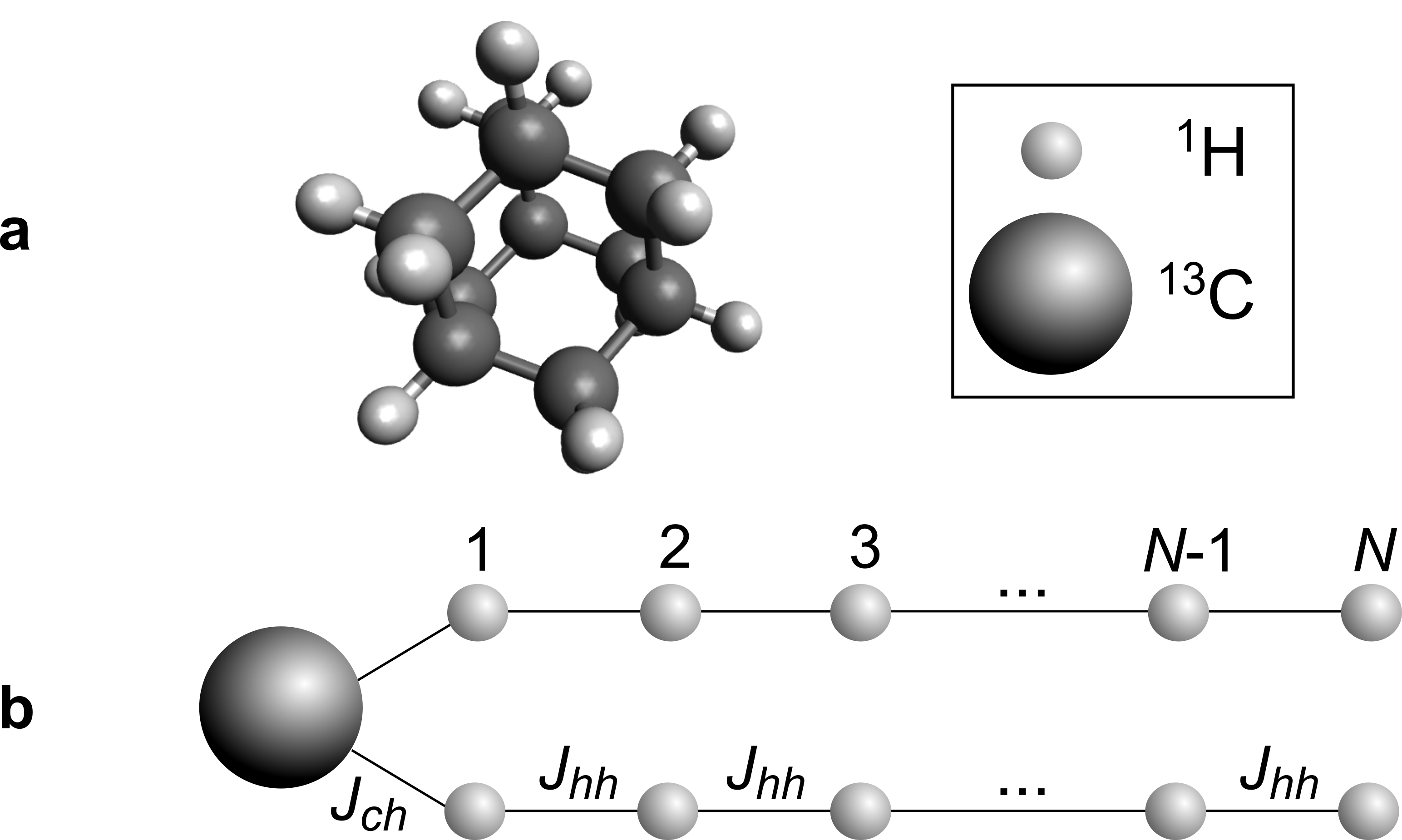}\caption{Spin system: \textbf{a} Adamantane molecule and \textbf{b} representation
of our spin system model composed by a$^{13}C$ atom coupled to two
linear chains of $N$ $^{1}H$ atoms each.\label{fig:sistema}}
\end{figure}

\textit{Theoretical model:} Our model is inspired on the adamantane
molecule ($C_{10}H_{16}$), Figure \ref{fig:sistema}\textbf{a}, which
is composed of six $CH_{2}$ groups and four $CH$ groups. The $^{13}C$
nuclear spins ($S=1/2$) are approximately 1.1\% of all the carbon
spins contained in an adamantane sample. In this case we can disregard
the carbon-carbon interaction and, although a carbon spin couples
with several $^{1}H$ spins close to it, the coupling strength is
inversely proportional to the distance between the respective spins,
which allow us to keep only the first-neighbor interactions. Therefore,
we can consider each $^{13}C$ nucleus as an independent spin surrounded
by a large number of hydrogen nuclear spins ($I=1/2$) that acts as
a bath to the carbons \citep{Alvarez2010,Ajoy2011,Souza2011}.

The experiments were performed at room temperature with a sample of
polycrystalline adamantane subjected to a high intensity magnetic
field using a Varian 500 MHz Nuclear Magnetic Resonance (NMR) spectrometer.
We consider $\boldsymbol{S}=S_{x}\boldsymbol{i}+S_{y}\boldsymbol{j}+S_{z}\boldsymbol{k}$
and $\boldsymbol{I=}I_{x}\boldsymbol{i}+I_{y}\boldsymbol{j}+I_{z}\boldsymbol{k}$
as the nuclear spin operators for $^{13}C$ and $^{1}H$, respectively,
with $S_{\alpha}$and $I_{\alpha}$, $\alpha=x,y,z$, the Pauli matrices.
In polycrystalline adamantane the spin-spin interaction is dominated
by the dipolar interaction \citep{Abragam1961,SlichterLivro1990},
while the interaction of the spins with the static magnetic field
results in a Zeeman splitting with angular frequency $\omega_{S}$
for the $^{13}C$ and $\omega_{I}$ for the $^{1}H$. Using the secular
approximation, we can neglect the terms of the dipolar coupling Hamiltonian
that do not commute with the strong Zeeman interaction \citep{Abragam1961}.
For modeling the entire spin system, here we propose a configuration
of two linear spin chains, as illustrated in Fig. \ref{fig:sistema}\textbf{b}.
We consider that the $^{13}C$ nucleus is coupled only to the first
hydrogen of each array and each hydrogen nucleus is coupled only to
its first neighbors. The total Hamiltonian for this model written
in the rotating frame at the Larmor frequencies is of the form \citep{SlichterLivro1990}
$H=H_{SE}+H_{E}$, where $H_{SE}\text{ and }H_{E}$ are, respectively,
the Hamiltonians of the system-environment interaction and of the
environment, such that 

\begin{alignat}{1}
{\displaystyle H_{SE}}= & J_{ch}\sum_{\alpha=a,b}S_{z}I_{z}^{\alpha,1},\label{eq:H-SE}\\
H_{E}= & J_{hh}\sum_{\alpha=a,b}\sum_{k=1}^{N-1}\left[2I_{z}^{\alpha,k}I_{z}^{\alpha,k+1}-\left(I_{x}^{\alpha,k}I_{x}^{\alpha,k+1}+I_{y}^{\alpha,k}I_{y}^{\alpha,k+1}\right)\right],\label{eq:H-E}
\end{alignat}
where the index $\alpha=a,b$ represents the arrays of hydrogen spins,
being the first one of each array coupled directly to the $^{13}C$.
The index \textit{$k$} represents the $k^{th}$ $^{1}H$ of the bath,
being \textit{$2N$} the total number of $^{1}H$ nuclear spins in
the bath system ($N$ in each array). $J_{ch}$ and $J_{hh}$ are,
respectively, the carbon-hydrogen and hydrogen-hydrogen coupling constants.

The natural interaction between the carbon and the hydrogen is not
able to promote flips on the qubit system (carbon spin), thus not
being immediately useful for our proposal. To engineer the desired
reservoir we firstly need to build up interactions that allow the
exchange of energy between the system and the environment. To this
end, we need to manipulate the nuclear spins in order to obtain an
effective suitable Hamiltonian. This can be done by applying the sequence
of radiofrequency (r.f.) pulses, being each one followed by a free
evolution governed by the system Hamiltonian $H$ during a short time
interval $\Delta t$. This sequence is then described by the evolution
operator

\begin{equation}
U(\tau_{c})=e^{-\frac{i}{2\hbar}H\Delta t}P_{4}e^{-\frac{i}{\hbar}H\Delta t}P_{3}e^{-\frac{i}{\hbar}H\Delta t}P_{2}e^{-\frac{i}{\hbar}H\Delta t}P_{1}e^{-\frac{i}{2\hbar}H\Delta t},\label{eq:U-total-1}
\end{equation}
where $P_{1}=exp\left[-i\frac{\pi}{2}\left(S_{x}+I_{x}\right)\right]$,
$P_{2}=exp\left[i\frac{\pi}{2}\left(S_{x}+I_{x}\right)\right]$, $P_{3}=exp\left[-i\frac{\pi}{2}\left(S_{y}+I_{y}\right)\right]$,
and $P_{4}=exp\left[i\frac{\pi}{2}\left(S_{y}+I_{y}\right)\right]$
are r.f. pulse operators that make the spins \textit{$I$} and \textit{$S$}
to flip at angles $\nicefrac{\pi}{2}$ in the $x$, $-x$, $y$ and
$-y$ directions, respectively.

\begin{figure}[h]
\begin{centering}
\includegraphics[width=1\columnwidth]{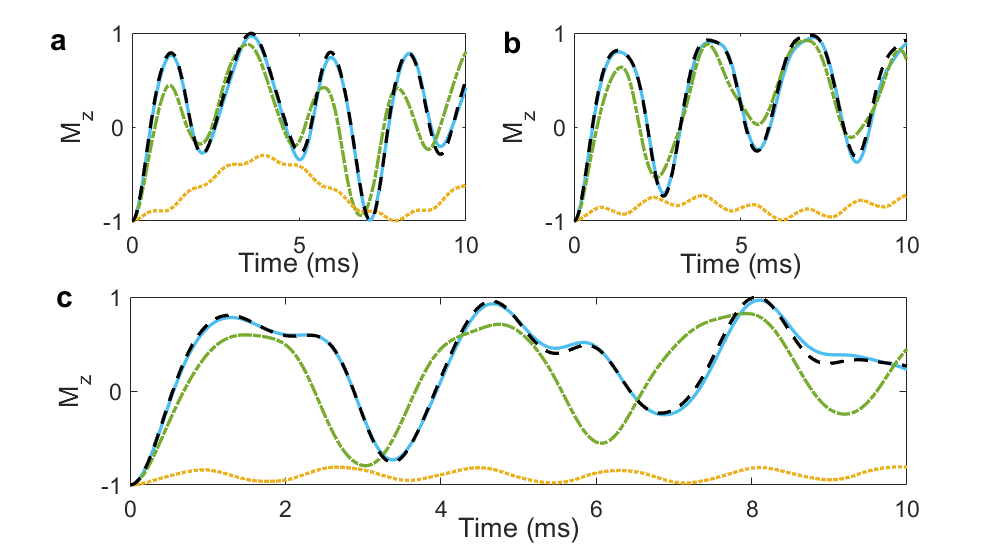}
\par\end{centering}
\caption{\textcolor{black}{Time evolution of the }\textit{\textcolor{black}{$z$}}\textcolor{black}{{}
component of the magnetization of the carbon nuclear spin ($M_{z}$)
coupled to two symmetric chains containing a total of }\textbf{\textcolor{black}{a}}\textcolor{black}{{}
6, }\textbf{\textcolor{black}{b}}\textcolor{black}{{} 8 and }\textbf{\textcolor{black}{c}}\textcolor{black}{{}
10 spins of hydrogen. The dashed black line was derived using $H_{eff}$
while the yellow (dotted), green (dashed-dotted) and blue (full) lines
refer to the simulations of $100$ (}$\Delta t=15.10$ $\mu s$\textcolor{black}{),
225 (}$\Delta t=1.228$ $\mu s$\textcolor{black}{) and 250 (}$\Delta t=0.10$
$\mu s$\textcolor{black}{) cycles, respectively, with }pulse sequence
described by equation \ref{eq:U-total-1}\textcolor{black}{. }The
parameters used are: $\tau_{p}=9.89$ $\mu s$, $J_{ch}^{eff}=550$
rad $s^{-1}$, $J_{hh}^{eff}=980$ rad $s^{-1}$. The initial state
is $|1\rangle$ for the carbon nuclear spin and $|0\rangle$ for all
hydrogen spins.\textcolor{black}{\label{fig:variando-delta-t}}}
\end{figure}

We can rewrite the equation (\ref{eq:U-total-1}) as a single evolution
operator governed by an effective Hamiltonian, i.e., $U(t)=e^{-\frac{i}{\hbar}H_{eff}t}$,
which can be calculated using average Hamiltonian theory \citep{SlichterLivro1990}:
\begin{equation}
H_{eff}=\bar{H}_{0}+\bar{H}_{1}+\bar{H}_{2}+...\bar{H}_{p}.
\end{equation}
The terms of $H_{eff}$ can be calculated from the Magnus' expansion
\citep{Magnus1954}, where $\bar{H}_{0}$ is the first approximation
for the Hamiltonian\textbf{ }and $\bar{H}_{1}\text{, }\bar{H}_{2},...,\bar{H}_{p}$
are the correction terms \citep{Maricq1982}. The time interval between
the different pulses used in this work is designed to be small enough
such that the correction terms are reduced close to zero, remaining\textbf{
}only the zeroth-order term $\bar{H}_{0}$, which is given by \citep{SlichterLivro1990}

\begin{equation}
\bar{H}_{0}=\frac{1}{t}\sum_{j=1}^{M}\Delta t_{j}\tilde{H}_{j},
\end{equation}
where $\tilde{H}_{j}=T_{j}^{\dagger}HT_{j}$, ${\displaystyle T_{j}=\prod_{m=1}^{M}P_{m}}$,
and $t={\displaystyle \sum_{j=1}^{M}}\Delta t_{j}$, being \textit{$M$
}the number of r.f. pulses. \textbf{$\Delta t_{j}$} is the time duration
that the system evolves under the $\tilde{H}_{j}$ Hamiltonian. As
a result, considering our linear spin chain model described by the
Hamiltonian $H$ (\ref{eq:H-E}), we derive the effective Hamiltonian
for the carbon-hydrogen and hydrogen-hydrogen interactions:

\begin{alignat}{1}
{\displaystyle H_{eff}^{SE}=} & J_{ch}^{eff}\sum_{\alpha=a,b}\left(2S_{z}I_{z}^{\alpha,1}+S_{x}I_{x}^{\alpha,1}+S_{y}I_{y}^{\alpha,1}\right),\label{eq:HSE_efetivo}\\
H_{eff}^{E}= & J_{hh}^{eff}\sum_{\alpha=a,b}\sum_{k=1}^{N-1}\left[2I_{z}^{\alpha,k}I_{z}^{\alpha,k+1}-\left(I_{x}^{\alpha,k}I_{x}^{\alpha,k+1}+I_{y}^{\alpha,k}I_{y}^{\alpha,k+1}\right)\right].\label{eq:HE_efetivo}
\end{alignat}
$J_{ch}^{eff}=J_{ch}/4\text{ and }J_{hh}^{eff}=J_{hh}/4$\textbf{
}are, respectively, the effective coupling constants of the interactions
between the nuclear spin of the $^{13}C$ and its first neighboring
hydrogens and between the nuclear hydrogen spins of each chain obtained
after the application of the pulse sequence described in equation
(\ref{eq:U-total-1}). In Figure \ref{fig:variando-delta-t} we plot
the magnetization of the carbon spin in $z$ direction ($M_{z}=\left\langle S_{z}\right\rangle $)
as a function of time, derived by solving numerically the Schrödinger
equation, either using the effective Hamiltonian (dashed line), Eqs.
(\ref{eq:HSE_efetivo}) and (\ref{eq:HE_efetivo}), or by simulating
the propagator (\ref{eq:U-total-1}) (full, dotted and dashed-dotted
lines). We considered the initial state of the system as $|1\rangle$
(excited) for the carbon nuclear spin and $|0\rangle$ (ground) for
all hydrogen nuclear spins. The time duration of each pulse was fixed
as $\tau_{p}=9.89$ $\mu s$ in all cases. However, we considered
different time intervals between each pulse: $\Delta t=15.10$ $\mu s$,
$1.22$ $\mu s$ and $0.10$ $\mu s$, for a sequence of $100$, $225$
and $250$ cycles during the total time evolution of $10$ $ms$.
We can observe that the lower the value of $\Delta t$, the better
the match between the results predicted via $H_{eff}$ and those obtained
from the pulse sequence described by equation \ref{eq:U-total-1}.

\begin{figure}[h]
\begin{centering}
\includegraphics[width=0.9\columnwidth]{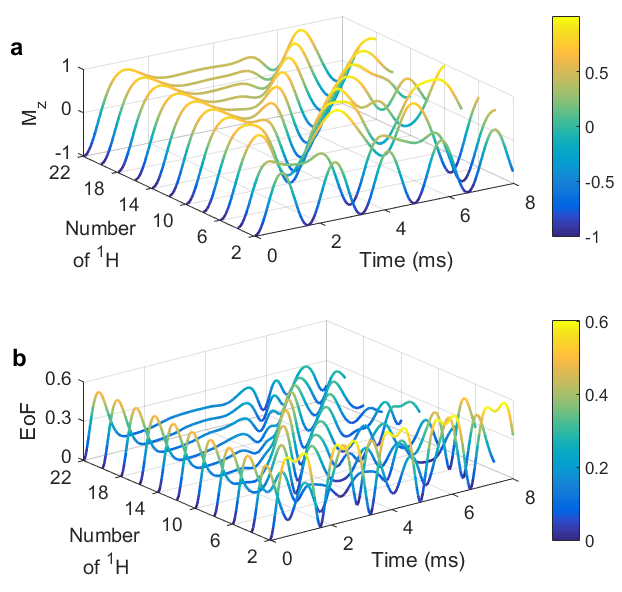}
\par\end{centering}
\caption{\textbf{a}\textit{ $z$} component of the magnetization of the carbon
nuclear spin as a function of time and number of qubits of the bath,
derived via solution of the Schrödinger equation with $H_{eff}$ and
\textbf{b} $EoF$ between the carbon and first hydrogen spin (first
array). The parameters used are: $\tau_{p}=9.89$ $\mu s$, $\Delta t=0.10$
$\mu s$, $J_{ch}^{eff}=550$ rad $s^{-1}$ and $J_{hh}^{eff}=980$
rad $s^{-1}$. The initial state is $|1\rangle$ for the carbon nuclear
spin and $|0\rangle$ for all hydrogen spins.\textbf{ }\label{fig:variandoN}}
\end{figure}

\begin{figure}[h]
\begin{centering}
\includegraphics[width=0.9\columnwidth]{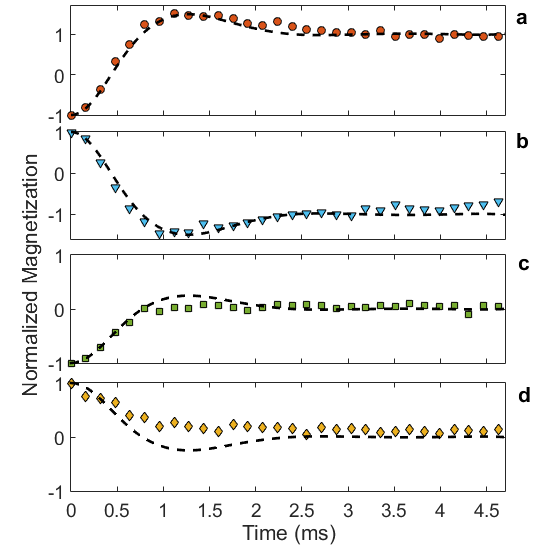} 
\par\end{centering}
\caption{Normalized $z$ component of the magnetization of the carbon nuclear
spin ($M_{z}$) as a function of time for the initial states: \textbf{a}
Carbon in $|1\rangle$ and hydrogens in $|0\rangle$; \textbf{b} Carbon
in $|0\rangle$ and hydrogens in $|1\rangle$; Hydrogens in $\rho=\frac{1}{2}\left(|0\rangle\langle0|+|1\rangle\langle1|\right)$
and carbon in \textbf{c} $|1\rangle$ and \textbf{d} $|0\rangle$.
The symbols refer to experimental data while the dashed lines are
derived theoretically via $H_{eff}$ (with $N=22$ hydrogen spins).
The parameters used here are $J_{ch}^{eff}=550$ rad $s^{-1}$and
$J_{hh}^{eff}=980$ rad $s^{-1}$.\label{fig:banho_PosNeg}}
\end{figure}

The control of the spin temperature is done by adjusting the hydrogen
nuclear spin state \citep{Purcell1951,Quan2007,Carr2013}. For example,
by inverting the population of the hydrogen spins, the environment
can be treated as having effective negative temperatures \citep{Struchtrup2018}.
To study the dynamics and thermalization of the carbon spin system
we consider different sizes of the environment, for different initial
states. Fig. \ref{fig:variandoN}\textbf{a} shows how the dynamics
of our carbon spin system behaves when we increase the size of the
environment (number of hydrogen spins) when the carbon spin system
is prepared in the excited state $|1\rangle$ and the environment
spins with positive temperature, i.e., all of them in the ground state
$|0\rangle$. The coupling strengths $J_{ch}^{eff}=550$ rad $s^{-1}$and
$J_{hh}^{eff}=980$ rad $s^{-1}$ were calibrated from the experimental
data shown in Fig. \textcolor{magenta}{\ref{fig:banho_PosNeg}}. We
observe that the greater the number of hydrogens in the spin system,
the better the thermalization. In Fig. \ref{fig:variandoN}\textbf{b}
we also plot the degree of entanglement between the carbon and the
first hydrogen spin (first array), quantified by the Entanglement
of Formation ($EoF$) \citep{Wootters1998}. A general bipartite system
is entangled if the global density matrix of the composite system
$\rho$ can not be written as as separable state, i.e., if $\rho\ne\sum_{i}P_{i}\rho_{C}^{(i)}\otimes\rho_{H}^{(i)}$,
being $\rho_{C}^{(i)}$ ($\rho_{H}^{(i)}$) any possible reduced density
matrix for the carbon (hydrogen) nuclear spin. When a system is in
a separable (not entangled) state $EoF=0$, while $EoF=1$ for maximally
entangled states. From Fig. \ref{fig:variandoN}\textbf{b} we see
that, for chains of few hydrogen spins, the entanglement between carbon
and the first hydrogen spin oscillates all the time. But, as we increase
the number of hydrogen spins, the carbon spin initially gets highly
entangled with the first hydrogen spin, and then the degree of entanglement
decreases, becoming close to zero. This means that entanglement moves
to the other hydrogen spins, disentangling the carbon spin from the
hydrogen chains. At the same time, the state of the carbon spin approaches
the initial hydrogen spin state. This is a clear signature that the
hydrogen spin chains work out as a real thermal bath for the carbon
spin, as desired.

\textit{Experimental results:} We prepared different initial states
for $^{13}C$ and $^{1}H$ simulating different temperatures of the
spin bath for the main qubit (carbon spin) and measured the expected
value of $z$ component of magnetization. Initially we start from
the thermal equilibrium at room temperature. In this high-temperature
limit, the spin states are almost equally populated with the excess
of the lower energy state on the order of $10^{-5}$. Firstly the
excess population of the carbon spins is prepared in the excited state
$|1\rangle$ and the hydrogen spins are kept in the thermal equilibrium
state (with excess of population in the ground state $|0\rangle$),
Figure \ref{fig:banho_PosNeg}\textbf{a}. Then we prepared the excess
population of carbon spins in the state $|0\rangle$ and the hydrogen
spins in the thermal equilibrium with negative temperature (with excess
of population in the state $|1\rangle$), Figures \ref{fig:banho_PosNeg}\textbf{b}.
This is obtained by inverting the population of all hydrogen spins
with r.f. pulses. During a short period of time, compared to the thermal
relaxation time of the sample, the transitions between the magnetic
energy levels can be neglected, keeping the hydrogen spins in an state
corresponding to the negative room temperature thermal state. By modulating
the system-environment interaction, i.e., the carbon-hydrogen interaction,
with the sequence of r.f. pulses it is possible to promote energy
exchange between the system and the environment. The resulting situation
is equivalent to put the carbon spins in contact to a thermal bath
with negative temperature. Still in Figure \ref{fig:banho_PosNeg},
we do the same for the bath at infinite temperature, i.e. the hydrogen
spins of the bath is prepared in the mixed state without population
excess $\rho=\frac{1}{2}\left(|0\rangle\langle0|+|1\rangle\langle1|\right)$,
Figures \ref{fig:banho_PosNeg}\textbf{c} and \ref{fig:banho_PosNeg}\textbf{d}.
In all cases we can observe, in very good agreement between the experimental
and the prediction of our model, that the qubit system tends to thermalize
in the state of the qubit bath, and this happens even when the qubit
system is in the ground state and the qubits bath are in the excited
state.

Slight difference between experimental and theoretical data can be
observed in different regions in all panels of Figure \ref{fig:banho_PosNeg}.
The origin of these errors are mainly due to r.f. pulse imperfections
and the natural relaxation of the spins. In NMR, the thermal relaxation
time is associated to the spin-lattice relaxation, which occurs at
a characteristic times $T_{1}$, which for our system are $T_{1}^{H}=0.9$
$s$ and $T_{1}^{C}=1.6$ $s$. These source of errors may reduce
the fidelity $\mathcal{F}$ \citep{Wang2008}, which is a parameter
commonly used to quantify the performance of quantum operations by
checking the compatibility between experimental and theoretical operators.
We can map a process matrix into a set of transformations of unitary
evolutions from the initial and final density matrices \citep{Chuang1997}.
For this purpose we use the equation $\rho_{f}={\displaystyle \sum_{mn}}\chi_{mn}E_{m}\rho_{i}E_{n}^{\dagger}$,
where $\rho_{i}$ and $\rho_{f}$ are the density matrices at the
beginning and end of the process, and the operators $E_{m}=I,S_{x},iS_{y},S_{z}$
must form a basis. Thus the propagator $\chi_{mn}$ for the process
can therefore be quantified.

\begin{figure}[H]
\begin{centering}
\includegraphics[width=1\columnwidth]{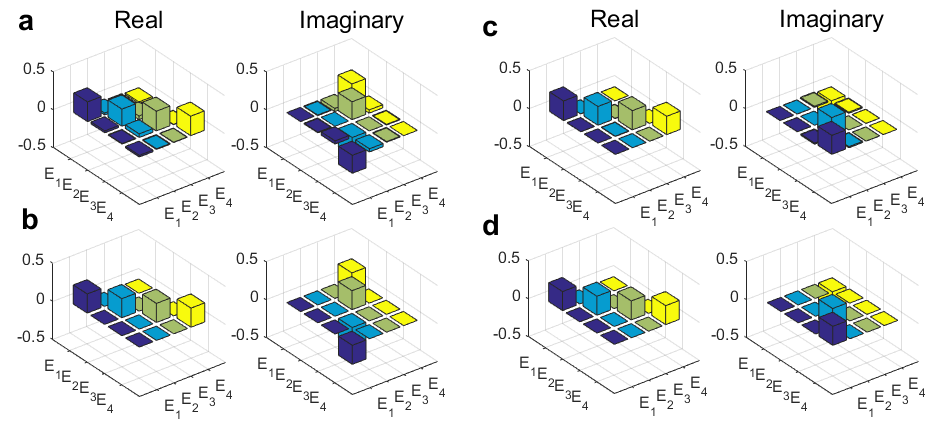} 
\par\end{centering}
\caption{Theoretical and experimental process matrices of $\chi_{mn}$ of the
thermalization processes of the Figures \ref{fig:banho_PosNeg}\textbf{a}
and \ref{fig:banho_PosNeg}\textbf{b}. The Figures in the left (right)
are the real (imaginary) part. \textbf{a} and \textbf{b} are, respectively,
the experimental and theoretical process matrices of the Fig. \ref{fig:banho_PosNeg}\textbf{a}.
Panels \textbf{c} and \textbf{d} are, respectively, the experimental
and theoretical process matrices of the \ref{fig:banho_PosNeg}\textbf{b}.
\label{fig:matriz de processo} }
\end{figure}

We performed a quantum tomography process to obtain the implemented
propagator and use $\mathcal{F}$ to compare the theoretical and experimental
results for $\chi_{mn}$. The fidelity $\mathcal{F}$ between the
experimental and ideal matrices of the thermalization process at positive
temperature, Figures \ref{fig:matriz de processo}\textbf{a} and \ref{fig:matriz de processo}\textbf{b},
and of the thermalization process at negative temperature, Figures
\ref{fig:matriz de processo}\textbf{c} and \ref{fig:matriz de processo}\textbf{d},
are 0.999 and 0.984, respectively.

\textit{Quantum thermal machine of a single reservoir:} As an application
for this reservoir technique here implemented, we propose a single
reservoir quantum thermal machine at an effective negative temperature.
The spin 1/2 of the $^{13}C$ nucleus is the working medium, and the
ensemble of spin 1/2 of $^{1}H$ nuclei plays the role of the hot
thermal reservoir. The Quantum thermal machine can be described by
the relation

\begin{equation}
\rho_{ground}\overset{thermalization}{\rightarrow}\rho_{1}\overset{U(\tau)}{\rightarrow}\rho_{2}\overset{thermalization}{\rightarrow}\rho_{1}.\label{eq:etapas_ciclo}
\end{equation}

Firstly the qubit system (carbon nucleus) is put in contact with the
hot reservoir, to absorb heat from it (step described in Fig. \ref{fig:banho_PosNeg}\textbf{b},
i.e., step $\rho_{ground}\rightarrow\rho_{1}$ in relation (\ref{eq:etapas_ciclo})).
Therefore, the $^{13}C$ spin will initially be in a thermal state
equivalent to $\rho_{1}=e^{-\beta_{1}H_{1}}/Z_{1}$, where $H_{1}=-\frac{1}{2}\hbar\omega_{1}S_{z}$
is Zeeman Hamiltonian, being $\hbar$ the Planck's constant, $\omega_{1}$
the Larmor frequency of the $^{13}C$ nuclear spin, $Z_{1}$ the partition
function, and $\beta_{1}=1/k_{B}T$ such that $T<0$ is the spin temperature
and $k_{B}$ is the Boltzmann constant. Then, our thermodynamic cycle
consists of two steps. (\textit{i}) In the first step an unitary evolution
operator $U(\tau)$ is applied to bring the $^{13}C$ spin state to
$\rho_{2}=U(\tau)\rho_{1}U^{\dagger}(\tau)$ (step $\rho_{1}\rightarrow\rho_{2}$
in relation (\ref{eq:etapas_ciclo})). The Hamiltonian $H_{1}$ changes
in this process, which is the one where we can extract work from the
machine, then turning to its initial form at the end of this step.
(\textit{ii}) The second step consists of thermalization with the
hot thermal reservoir (step $\rho_{2}\rightarrow\rho_{1}$ in relation
(\ref{eq:etapas_ciclo})), when the qubit absorbs heat and then goes
back to the state $\rho_{1}$. Here the Hamiltonian $H_{1}$ remains
unchanged.

\begin{figure}
\includegraphics[width=1\columnwidth]{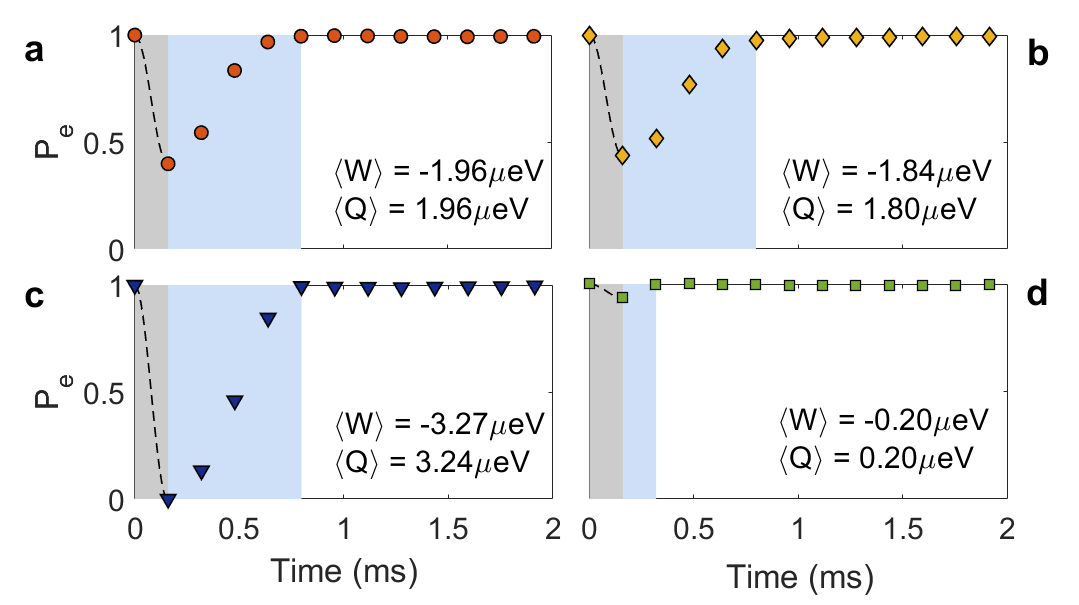}

\caption{Population ($\mathrm{P_{e}}$) of the excited state as a function
of time for different unitary operations. We consider the beginning
of our cycle in $t=0$, starting from a thermalized state with the
environment at a negative effective temperature. The unitary operations
performed were: \textbf{a} $U_{x}=exp\left(-i\frac{\pi}{2}S_{x}\right)$,
\textbf{b} $U_{y}=exp\left(-i\frac{\pi}{2}S_{y}\right)$, \textbf{c}
$U_{\pi}=exp\left(-i\pi S_{x}\right)$ and \textbf{d} $U_{I}=exp\left(-2i\pi S_{x}\right)$.
The work is calculated in the unitary operation region (gray region)
and the heat exchanged with the hot thermal reservoir is calculated
in the thermalization region with the reservoir (blue region).\label{fig:maquina}}
\end{figure}

In Figure \ref{fig:maquina} we show the application of four different
unitary operators: Fig. \ref{fig:maquina}\textbf{a} $U_{x}=exp\left(-i\frac{\pi}{2}S_{x}\right)$,
Fig. \ref{fig:maquina}\textbf{b} $U_{y}=exp\left(-i\frac{\pi}{2}S_{y}\right)$,
Fig. \ref{fig:maquina}\textbf{c} $U_{\pi}=exp\left(-i\pi S_{x}\right)$,
and Fig. \ref{fig:maquina}\textbf{d} $U_{I}=exp\left(-2i\pi S_{y}\right)=\mathbb{I}$,
where $U_{x}$ and $U_{y}$ are operators that represent pulses in
the $x$ and $y$ directions, allowing the $^{13}C$ spin to flip
by an angle of $\nicefrac{\pi}{2}$. $U_{\pi}$ is the operator which
rotates the spin by an angle of $\pi$ and $U_{I}$ is an identity
operator (a complete rotation around the Bloch sphere). The total
magnetization in the $x$, $y$ and $z$ directions - $M_{U}=(M_{x},M_{y},M_{z})$
- immediately after each unitary operation (second point in Fig. \ref{fig:maquina})
is $M_{U_{x}}=(0.11,-0.92,0.20)\mathrm{a.u.}$, $M_{U_{y}}=(0.96,-0.05,0.13)\mathrm{a.u.}$,
$M_{U_{\pi}}=(-0.01,-0.01,1.0)\mathrm{a.u.}$, and $M_{U_{I}}=(-0.01,0.15,-0.87)\mathrm{a.u.}$.

The gray region in Fig. \ref{eq:etapas_ciclo} refers to the carbon
nucleus leaving the thermalized state with the hot reservoir (spin
bath) \textbf{$\rho_{1}$} and going to the state \textbf{$\rho_{2}$}
as a result of the unitary operation. The average work $\left\langle W\right\rangle $
performed by the quantum thermal machine is then given by (see SM
of \citep{deAssis2019})

\begin{equation}
\left\langle W\right\rangle =Tr\left[\rho_{2}H_{1}\right]-Tr\left[\rho_{1}H_{1}\right]=-\hbar\omega_{1}\xi\tanh\left({\textstyle \frac{1}{2}}\left|\beta_{1}\right|\hbar\omega_{1}\right),
\end{equation}
where $\xi=\left|\langle1\vert U(t)\vert0\rangle\right|^{2}\ge0$
is the transition probability between initial state $\vert0\rangle$
and $\vert1\rangle$ and $\beta_{1}=-\left|\beta_{1}\right|<0$. Here
$\left\langle W\right\rangle <0$ means that the work is done by the
machine. The blue region in Fig. \ref{fig:maquina} refers to the
thermalization process with the hot reservoir, immediately after the
carbon spin being in the state \textbf{$\rho_{2}$}, which brings
it back to the state $\rho_{1}.$ The average heat $\left\langle Q\right\rangle $
exchanged with the hot thermal reservoir is then (see SM of \citep{deAssis2019})

\begin{equation}
\left\langle Q\right\rangle =Tr\left[\rho_{1}H_{1}\right]-Tr\left[\rho_{2}H_{1}\right]=\hbar\omega_{1}\xi\tanh\left({\textstyle \frac{1}{2}}\left|\beta_{1}\right|\hbar\omega_{1}\right).
\end{equation}

The results obtained experimentally for the different unitary operations
are: $\left\langle W\right\rangle _{U_{x}}=-\left(1.97\pm0.28\right)\mu eV$,
$\left\langle W\right\rangle _{U_{y}}=-\left(1.84\pm0.21\right)\mu eV$,
$\left\langle W\right\rangle _{U_{\pi}}=-\left(3.27\pm0.22\right)\mu eV$,
and $\left\langle W\right\rangle _{U_{I}}=-\left(0.20\pm0.21\right)\mu eV$.
The heat absorbed from the reservoir are: $\left\langle Q\right\rangle _{U_{x}}=\left(1.96\pm0.29\right)\mu eV$,
$\left\langle Q\right\rangle _{U_{y}}=\left(1.79\pm0.18\right)\mu eV$,
$\left\langle Q\right\rangle _{U_{\pi}}=\left(3.24\pm0.22\right)\mu eV$,
and $\left\langle Q\right\rangle _{U_{I}}=\left(0.20\pm0.21\right)\mu eV$.
Thus, the efficiency $\eta=\left|\frac{\left\langle W\right\rangle }{\left\langle Q\right\rangle }\right|$
for each process is $\eta_{U_{x}}=1.00\pm0.05$, $\eta_{U_{y}}=1.03\pm0.06$,
$\eta_{U_{\pi}}=1.01\pm0.02$ and $\eta{}_{U_{I}}=1.00\pm0.12$, i.e.,
whenever the final state is different from the initial one, the machine
absorbs heat and converts it entirely into liquid work.

To conclude, here we have applied the reservoir engineering technique
to build up reservoirs with arbitrary temperatures, even effective
negative ones, for qubit systems. Our system is composed by a nuclear
carbon spin (the main qubit) coupled to a large number of nuclear
hydrogen spins. By properly manipulating the initial hydrogen state
and the carbon-hydrogen and hydrogen-hydrogen nuclear spin interactions,
the effective dynamics describes the interaction of a qubit with reservoirs
at arbitrary temperatures. We have shown theoretically that, the larger
the number of hydrogen spins, the better the hydrogen arrays work
out as a bath for the main qubit, which is in excellent agreement
with our experimental results. As an application, we have implemented
a single reservoir heat engine. Several papers have discussed how
to increase the efficiency of quantum thermal machines, for example
via the use of quantum coherence \citep{Scully2003,Serra2019a} or
artificial environments as squeezed reservoirs \citep{Myatt2000,Murch2013}.
However, as shown in our work, the simple use of a single reservoir
with negative effective temperature allows us to obtain maximum efficiency
($\eta$ = 1) regardless of the unitary transformation performed on
the qubit (making sure that the final state is different from the
initial one), thus being, to the best of our knowledge, the simpler
and most efficient quantum thermal machine implemented so far. Naturally,
to achieve this result we need to build up this artificial reservoir,
with inverted population, which demands energy, but this is the price
we have to pay for this new technology. For instance, this is quite
similar to what happens to laser, another system based on inverted
population, which is responsible for great scientific and technological
advances in the last decades. Thus, we believe our single reservoir
heat engine can be another interesting application for inverted population
systems and that the results presented here can be very useful for
investigating fundamental and applications of quantum thermodynamics
in general, for instance, to study quantum thermal engines which require
specific kinds of reservoirs \citep{Klaers2017,deAssis2019,deAssis2019b}.
\begin{acknowledgments}
This work was supported by the Brazilian National Institute of Science
and Technology for Quantum Information (INCT-IQ) Grant No. 465469/2014-0
and by the Coordenação de Aperfeiçoamento de Pessoal de Nível Superior
- Brasil (CAPES) - Finance Code 001. C.J.V.-B. also thanks the support
by the São Paulo Research Foundation (FAPESP) Grants No. 2013/04162-5
and 2019/11999-5, and the National Council for Scientific and Technological
Development (CNPq) Grant No. 307077/2018-7. A. M. S. acknowledges
support from the Brazilian agencies FAPERJ (Grant No. 203.166/2017)
and CNPq (Grant No. 304986/2016-0). N.G.A. also thanks the support
by the FAPEG agency.
\end{acknowledgments}

\bibliographystyle{unsrt}
\bibliography{report,referencias}

\begin{thebibliography}{10}

\bibitem{Gemmer2004}
J.~Gemmer, M.~Michel, and G.~Mahler.
\newblock {\em Quantum Thermodynamics}.
\newblock Springer Verlag, Berlin, 2004.

\bibitem{Alicki1979}
R.~Alicki.
\newblock The quantum open system as a model of the heat engine.
\newblock {\em Journal of Physics A: Mathematical and General},
  12(5):L103--L107, may 1979.

\bibitem{Quan2007}
H.~T. Quan, Yu-xi Liu, C.~P. Sun, and Franco Nori.
\newblock Quantum thermodynamic cycles and quantum heat engines.
\newblock {\em Phys. Rev. E}, 76:031105, Sep 2007.

\bibitem{Purcell1951}
E.~M. Purcell and R.~V. Pound.
\newblock A nuclear spin system at negative temperature.
\newblock {\em Phys. Rev.}, 81:279--280, Jan 1951.

\bibitem{Ramsey1956}
N.~F. Ramsey.
\newblock Thermodynamics and statistical mechanics at negative absolute
  temperatures.
\newblock {\em Phys. Rev.}, 103:20--28, Jul 1956.

\bibitem{Carr2013}
L.~D. Carr.
\newblock Negative temperatures?
\newblock {\em Science}, 339(6115):42--43, 2013.

\bibitem{Braun2013}
S.~Braun, J.~P. Ronzheimer, M.~Schreiber, S.~S. Hodgman, T.~Rom, I.~Bloch, and
  U.~Schneider.
\newblock Negative absolute temperature for motional degrees of freedom.
\newblock {\em Science}, 339(6115):52--55, 2013.

\bibitem{deAssis2019}
R.~J. de~Assis, T.~M. de~Mendon\c{c}a, C.~J. Villas-Boas, A.~M. de~Souza, R.~S.
  Sarthour, I.~S. Oliveira, and N.~G. de~Almeida.
\newblock Efficiency of a quantum otto heat engine operating under a reservoir
  at effective negative temperatures.
\newblock {\em Phys. Rev. Lett.}, 122:240602, Jun 2019.

\bibitem{Poyatos1996}
J.~F. Poyatos, J.~I. Cirac, and P.~Zoller.
\newblock Quantum reservoir engineering with laser cooled trapped ions.
\newblock {\em Phys. Rev. Lett.}, 77:4728--4731, Dec 1996.

\bibitem{Myatt2000}
C.~J. Myatt, B.~E. King, Q.~A. Turchette, C.~A. Sackett, D.~Kielpinski, W.~M.
  Itano, C.~Monroe, and D.~J. Wineland.
\newblock Decoherence of quantum superpositions through coupling to engineered
  reservoirs.
\newblock {\em Nature}, 403:269--273, Jan 2000.

\bibitem{Murch2013}
K.~W. Murch, S.~J. Weber, K.~M. Beck, E.~Ginossar, and I.~Siddiqi.
\newblock Reduction of the radiative decay of atomic coherence in squeezed
  vacuum.
\newblock {\em Nature}, 499:62--65, Jul 2013.

\bibitem{Carvalho2001}
A.~R.~R. Carvalho, P.~Milman, R.~L. de~Matos~Filho, and L.~Davidovich.
\newblock Decoherence, pointer engineering, and quantum state protection.
\newblock {\em Phys. Rev. Lett.}, 86:4988--4991, May 2001.

\bibitem{Werlang2008b}
T.~Werlang, R.~Guzm\'an, F.~O. Prado, and C.~J. Villas-B\^oas.
\newblock Generation of decoherence-free displaced squeezed states of radiation
  fields and a squeezed reservoir for atoms in cavity qed.
\newblock {\em Phys. Rev. A}, 78:033820, Sep 2008.

\bibitem{Pielawa2007}
S.~Pielawa, G.~Morigi, D.~Vitali, and L.~Davidovich.
\newblock Generation of einstein-podolsky-rosen-entangled radiation through an
  atomic reservoir.
\newblock {\em Phys. Rev. Lett.}, 98:240401, Jun 2007.

\bibitem{Werlang2008}
T.~Werlang and C.~J. Villas-Boas.
\newblock Theoretical method for the generation of a dark two-mode squeezed
  state of a trapped ion.
\newblock {\em Phys. Rev. A}, 77:065801, Jun 2008.

\bibitem{Prado2009}
F.~O. Prado, E.~I. Duzzioni, M.~H.~Y. Moussa, N.~G. de~Almeida, and C.~J.
  Villas-B\^oas.
\newblock Nonadiabatic coherent evolution of two-level systems under
  spontaneous decay.
\newblock {\em Phys. Rev. Lett.}, 102:073008, Feb 2009.

\bibitem{Verstraete2009}
F.~Verstraete, M.l~M. Wolf, and J.~Ignacio~Cirac.
\newblock Quantum computation and quantum-state engineering driven by
  dissipation.
\newblock {\em Nature Physics}, 5:633--636, Jul 2009.

\bibitem{Hama2018}
Yusuke Hama, William~J. Munro, and Kae Nemoto.
\newblock Relaxation to negative temperatures in double domain systems.
\newblock {\em Phys. Rev. Lett.}, 120:060403, Feb 2018.

\bibitem{Suter2011}
Gonzalo~A. \'Alvarez and Dieter Suter.
\newblock Measuring the spectrum of colored noise by dynamical decoupling.
\newblock {\em Phys. Rev. Lett.}, 107:230501, Nov 2011.

\bibitem{Suter2007}
Marko Lovri\ifmmode~\acute{c}\else \'{c}\fi{}, Hans~Georg Krojanski, and Dieter
  Suter.
\newblock Decoherence in large quantum registers under variable interaction
  with the environment.
\newblock {\em Phys. Rev. A}, 75:042305, Apr 2007.

\bibitem{Alvarez2010}
G.~A. Alvarez, A.~Ajoy, X.~Peng, and D.~Suter.
\newblock Performance comparison of dynamical decoupling sequences for a qubit
  in a rapidly fluctuating spin bath.
\newblock {\em Phys. Rev. A}, 82:042306, Oct 2010.

\bibitem{Ajoy2011}
A.~Ajoy, G.~A. Alvarez, and D.~Suter.
\newblock Optimal pulse spacing for dynamical decoupling in the presence of a
  purely dephasing spin bath.
\newblock {\em Phys. Rev. A}, 83:032303, Mar 2011.

\bibitem{Souza2011}
A.~M. Souza, G.A. Alvarez, and D.~Suter.
\newblock Robust dynamical decoupling for quantum computing and quantum memory.
\newblock {\em Phys. Rev. Lett.}, 106:240501, 2011.

\bibitem{Abragam1961}
A.~Abragam.
\newblock {\em Principles of magnetism}.
\newblock Oxford University Press, Oxford, 1962.

\bibitem{SlichterLivro1990}
C.~P. Slichter.
\newblock {\em Principles of magnetic resonance}, volume~3.
\newblock Springer, Berlin, 1990.

\bibitem{Magnus1954}
W.~Magnus.
\newblock On the exponential solution of differential equations for a linear
  operator.
\newblock {\em Communications on Pure and Applied Mathematics}, 7(4):649--673,
  1954.

\bibitem{Maricq1982}
M.~M. Maricq.
\newblock Application of average hamiltonian theory to the nmr of solids.
\newblock {\em Phys. Rev. B}, 25:6622--6632, Jun 1982.

\bibitem{Struchtrup2018}
H.~Struchtrup.
\newblock Work storage in states of apparent negative thermodynamic
  temperature.
\newblock {\em Phys. Rev. Lett.}, 120:250602, Jun 2018.

\bibitem{Wootters1998}
W.~K. Wootters.
\newblock Entanglement of formation of an arbitrary state of two qubits.
\newblock {\em Phys. Rev. Lett.}, 80:2245--2248, Mar 1998.

\bibitem{Wang2008}
X.~Wang, C.-S. Yu, and X.X. Yi.
\newblock An alternative quantum fidelity for mixed states of qudits.
\newblock {\em Physics Letters A}, 373(1):58 -- 60, 2008.

\bibitem{Chuang1997}
I.~L. Chuang and M.~A. Nielsen.
\newblock Prescription for experimental determination of the dynamics of a
  quantum black box.
\newblock {\em Journal of Modern Optics}, 44(11-12):2455--2467, 1997.

\bibitem{Scully2003}
Marlan~O. Scully, M.~Suhail Zubairy, Girish~S. Agarwal, and Herbert Walther.
\newblock Extracting work from a single heat bath via vanishing quantum
  coherence.
\newblock {\em Science}, 299(5608):862--864, 2003.

\bibitem{Serra2019a}
Patrice~A. Camati, Jonas F.~G. Santos, and Roberto~M. Serra.
\newblock Coherence effects in the performance of the quantum otto heat engine.
\newblock {\em Phys. Rev. A}, 99:062103, Jun 2019.

\bibitem{Klaers2017}
J.~Klaers, S.~Faelt, A.~Imamoglu, and E.~Togan.
\newblock Squeezed thermal reservoirs as a resource for a nanomechanical engine
  beyond the carnot limit.
\newblock {\em Phys. Rev. X}, 7:031044, Sep 2017.

\bibitem{deAssis2019b}
R.~J. de~Assis, C.~J. Villas-Boas, and N.~G. de~Almeida.
\newblock Feasible platform to study negative temperatures.
\newblock 52(6):065501, feb 2019.

\end{thebibliography}

\end{document}